\documentclass[12pt]{article}
\usepackage{amsmath}
\usepackage{epsfig}
\usepackage{amsfonts}
\usepackage{feynmp}

\title{"Isochronic" dynamical systems and nullification of amplitudes}
\author{Joanna Domienik\thanks{supported by the {\L}\'od\'z University grant No. 269},\\ 
 Piotr Kosi\'nski\thanks{supported by KBN grant No. 5 P03B 060 21} \\
Department of Theoretical Physics II \\
University of {\L}\'od\'z \\
Pomorska 149/153, 90 - 236 {\L}\'od\'z/Poland.}
\date{}

\begin{document}
\maketitle
\begin{abstract}
We construct the set of theories which share the property that the tree-level threshold amplitudes nullify even if both initial
and final states contain the same type of particles. The origin of this phenomenom lies in the fact that the reduced classical 
dynamics describes isochronic systems.
\end{abstract}

\newpage
\begin{fmffile}{fgraph}
The problem of multiparticle production has attracted much attention in the past decade \cite{b1}. It appeared that quite a 
detailed knowledge concerning the amplitudes of such processes is possible for special kinematics, in particular those 
involving particles produced at rest \cite{b2} $\div $\ \cite{b7}.

An interesting phenomenon that appeared here is the nullification of certain tree amplitudes at the threshold. For example,
for the process $2\rightarrow n$, with all final particles at rest, all amplitudes vanish except $n=2$\ and $n=4$\ in 
$ \Phi^4$\ unbroken theory and except $n=2$\ if $ \Phi \rightarrow -\Phi $\ symmetry is broken spontaneously \cite{b8}
$\div$\ \cite{b10}. Other theories were also analysed from this point of view and the nullification of tree $2\rightarrow n$\
amplitudes at the threshold has been discovered in the bosonic sector of electroweak model \cite{b11} and in the linear 
$\sigma$-model \cite{b12}. These results in general do not extend to the one-loop level \cite{b13}. One should also mention
that in more complicated theories the nullification takes place only provided some relations between parameters are satisfied \cite{b11}.
The origin of these relations (some hidden symmetry?) remains unclear and is obscured by the fact that nullification does not
survive, in general, beyond tree approximation.

In the very interesting papers Libanov, Rubakov and Troitsky \cite{b14}, \cite{b15} provided another example of threshold amplitudes
nullification in the tree approximation. They considered $\Phi^4$-theory with $O(2)$\ symmetry, the symmetry being softly
broken by the mass term. It appeared that the tree amplitudes describing the process of the production of $n_2$\ 
particles $\varphi_2$\ by $n_1$\ particles $\varphi_1$, all at rest, vanishes if $n_1$\ and $n_2$\ are coprime numbers
up to one common divisor $2$. Libanov et al. showed that the ultimate reason for nullification is that the $O(2)$-symmetry
survives, in some sense, when the symmetry breaking mass term is introduced. Let us sketch briefly their argument.
The starting point is the well-known fact that all Green functions in tree approximation are generated by the solution
of classical field equations with additional coupling to external sources and Feynman boundary conditions. Such a 
solution represents tree-graphs  contribution to one-point Green function in the presence of external sources. The consecutive
derivatives at vanishing sources provide the relevant Green functions in tree approximation. However, we can do even better
\cite{b4} (see also \cite{b16}). One considers the generating functional for the matrix elements of the field between 
the states containing arbitrary numbers of in- and out- on-shell particles. This functional can be obtained as follows 
\cite{b4}, \cite{b16}. Let the relevant Lagrangian be
\begin{eqnarray}
L(\Phi,\; \partial_{\mu}\Phi )=L_0(\Phi,\; \partial_{\mu}\Phi)+L_J(\Phi ), \label{w1}
\end{eqnarray}
where $\Phi \equiv (\Phi_i)$\ is the collection of fields, $L_0$\ contains all quadratic terms and $L_J$\ describes 
interactions. Consider the system of integral equations
\begin{eqnarray}
\Phi_i(x\mid \Phi_0)=\Phi_{0i}(x)+\int d^4y\Delta_{F_{ij}}(x-y)\frac{\partial L_{J}(\Phi )}{\partial \Phi_j(y)};\label{w2}
\end{eqnarray}
here $\Delta_{F_{ij}}$\ is the operator inverse to $\frac{\delta^2L_0}{\delta \Phi_i\delta \Phi_j}$\ with Feynman boundary conditions imposed and 
$ \Phi_{0i}(x)$\ is the combination, with arbitrary coefficients, of free-particle wave functions with positive (for incoming
particles) and negative (for outgoing particles) energies. Succesive derivatives with respect to these arbitrary coefficients 
give relevant matrix elements. Graphically, these matrix elements are given by sums of tree graphs with all external 
lines but one amputated and replaced by relevant wave functions. In order to obtain the corresponding $S$-matrix element
one has only to amputate the remaining propagator and go to mass shell with the corresponding fourmomentum. Eq. (\ref{w2})
implies
\begin{subequations}
\label{w3}
\begin{gather}
(\Box  \delta_{ij}+m^2_{ij})\Phi_j(x\mid \Phi_0)-\frac{\partial L_J}{\partial \Phi_i}
\mid_{\Phi_i\rightarrow \Phi_i(x\mid \Phi_0)}=0 \label{w3a}\\
\Phi_i(x\mid \Phi_0)\mid_{L_J=0}=\Phi_{0i}(x)\label{w3b}
\end{gather}
\end{subequations}
Things simplify considerably if all particles are at rest. All matrix elements become space-independent and only the time 
dependence remains to be determined. Eq. (\ref{w3}) is transformed to
\begin{eqnarray}
(\partial^2_t\delta_{ij}+m^2_{ij})\Phi_j(t\mid \Phi_0)-\frac{\partial L_J(\Phi )}{\partial \Phi_i}\mid_{\Phi_i=
\Phi_i(t\mid \Phi_0)}=0\label{w4}
\end{eqnarray}
We arrive at the set of nonlinear coupled oscillators. Tree expansion arises when we solve (\ref{w4}) pertubatively in
$L_J(\Phi )$. Libanov et al. have shown that nonvanishing amplitudes are produced if, in the course of solving (\ref{w4})
pertubatively, we are faced with the resonances. Then the solution diverges and this very divergence is cancelled when
the external line is amputated.

Divergent resonant solution means that we are looking for solution with diverging initial conditions. If, instead, we insist
on keeping initial conditions finite while aproaching resonance, the preexponential factor linear (in general-polynomical)
in time is produced. So, nonvanishing amplitudes are possible only if the expansion of $\Phi_i(t\mid \Phi_0)$,
eq.(\ref{w3}), in terms of
coupling constant$(s)$\ contains terms which are polynomial in time \cite{b16}. Libanov et al. have shown that, in the 
$O(2)$\ case, where the corresponding mechanical system in integrable, the symmetry related to the additional integral of motion 
prevents the resonances to appear. Consequently, the corresponding tree amplitudes vanish. 

Eventually, this nullification is a result of subtle cancellations of contributions coming from separate graphs. They can
be shown to result from Ward identities related to the above symmetry \cite{b17}.

Libanov et al. argued that the nullification described above should be valid in more general situation. Namely, the reduced
classical system, which describes tree amplitudes at the threshold, should exhibit a non-trivial symmetry with the property
that the infinitesimal transformation for at least one of the fields contains a term linear in this field or its derivative.
This conclusion can be supported by more detailed still simple arguments \cite{b18}.

One can understand the result of Ref. \cite{b15} from slightly different perspective \cite{b16}. Assume that the reduced
dynamical system (\ref{w4}) is integrable (and confining-this last requirement is, however, not crucial). Then one can 
introduce action-angle variables $(J_i,\;\theta_i)$\ and expand $\Phi_i(t\mid \Phi_0)$\ in multiple Fourier series
\begin{eqnarray}
\Phi_i(t\mid \Phi_0)=\sum\limits_{n_1,\;...,\;n_r}A_{i,\;n_1,\;...,\;n_r}(\underline{J},\;\underline{\lambda})
e^{i \sum\limits_{k=1}^{r} n_k\omega_k(\underline{J},\;\underline{\lambda})t};\label{w5}
\end{eqnarray}
here $\underline{\lambda}$\ stands for the set of coupling constants. As we have explained above, the resonancees are
related to the polynomial preexponential time dependence of separate terms in perturbative expansion. If one expands the 
righthand side of (\ref{w5}) in $\underline{\lambda}$\ such terms result from $\underline{\lambda}$-dependence of 
frequencies $\omega_k(\underline{J};\;\underline{\lambda})$. In general, $\omega_k(\underline{J},\;\underline{\lambda})$\
\underline{do} depend on $\underline{\lambda}$. However, with $\underline{J}=0$\ (under appropriate normalization of
$J$'s), $\omega_k(0;\;\underline{\lambda})$\ become the frequencies of harmonic part (i. e. the masses of particles) and 
do not depend on $\underline{\lambda}$. Now, the crucial point is that we are considering the amplitudes with different 
kinds of particles in incoming and outcoming states. Therefore, in the boundary condition (\ref{w3b}) we can put
\begin{eqnarray}
\Phi_{0i}=z_ie^{i\varepsilon_im_it},\;\varepsilon_i=\pm 1 \label{w6}
\end{eqnarray}
Then the nontrivial solutions with $\underline{J}=0$\ are possible (cf. the explicit solutions given in Ref. \cite{b14}),
i. e. the coefficients $A_{i;\;n_1,\;...,\;n_r}$\ are nonvanishing also for $\mid n_1 \mid +...+\mid n_r\mid \neq 0$. 
Eq. (\ref{w5}) takes the form
\begin{eqnarray}
\Phi_i(t\mid \Phi_0)=\sum\limits_{n_1,\;...,\;n_r}A_{i;\;n_1,\;...,\;n_r}(\underline{J};\;\underline{\lambda})
e^{i\sum\limits_{k=1}^rn_km_kt}\label{w7}
\end{eqnarray}
No terms polynomial in time appear in $\underline{\lambda}$\ expansion and the corresponding amplitudes do vanish.

In the above reasoning it is crucial that the boundary conditions take the form given by eq. (\ref{w6}). Such conditions 
admit the exact solutions corresponding to vanishing action variables. On the contrary, if the boundary conditions contain
the frequncies of both signs (which is unavoidable if both initial and final states contain the same particles) the solutions
with $\underline{J}=0$\ are exluded. This makes the problem whether the threshold amplitudes nullify more complicated.

We show below how one can construct field theories with vanishing threshold amplitudes (in the tree-graph approximation) with 
the same kind of particles both in initial and final states. The resulting theories are not renormalizable, yet they can be
viewed as low-energy effective theories in the sense of Weinberg \cite{b19}; 
moreover,  we are considering tree amplitudes only.

Assume that we have just one scalar field,
\begin{eqnarray}
L=\frac{1}{2}\partial_{\mu}\Phi \partial^{\mu}\Phi -V(\Phi), \label{w8}
\end{eqnarray}
so that the relevant amplitudes are $n\rightarrow n$\ with a single kind of particles in both states.

The reduced system has one degree of freedom so energy is the only time-independent integral of motion. The counterpart
of (\ref{w7}) reads
\begin{eqnarray}
\Phi(t\mid \Phi_0)=\sum\limits_nA_n(E,\;\underline{\lambda})e^{in\omega (E,\underline{\lambda})t}\label{w9}
\end{eqnarray}
Now, due to the fact that both initial and final states contain the same particles, $\Phi_0$\ must be the combination 
of both frequencies $\pm m$. The cross term produces nonzero contribution to the energy; so $E\neq 0$\ and 
$\omega (E,\; \underline{\lambda})$\ generalically depends on $\underline{\lambda}$. The only exception is the case
when $\omega (E,\; \underline{\lambda})$\ does not depend on 
$E,\;\omega (E,\; \underline{\lambda})=\omega (0,\; \underline{\lambda})\equiv m$. The general construction of systems with
the prescribed energy dependence of the frequency has been described in \cite{b20}. Recently, it has been applied \cite{b21}, 
\cite{b22} to the construction of certain superintegrable systems. The results of \cite{b21} and \cite{b22} imply the
following form of the
lagrangians describing trajectories with energy-independent frequency. Let 
$ \rho : {\bf R} \stackrel{onto}{\longrightarrow} {\bf R}$\
 be one-to-one and such 
that $ \rho \circ  \rho =id$. The relevant lagrangian reads ($\alpha >0$\ being an arbitrary parameter)
\begin{eqnarray}
L=\frac{1}{2}\dot{\Phi}^2-\alpha (\Phi -\rho (\Phi ))^2;\label{w10}
\end{eqnarray}
moreover, to get a nontrivial theory we must assume that $\rho$\ is decreasing. Also, without loss of generality we can take 
$\rho (0)=0$. The corresponding field theory reads
\begin{eqnarray}
L=\frac{1}{2}\partial_{\mu}\Phi \partial^{\mu}\Phi -\alpha (\Phi -\rho (\Phi ))^2 \label{w11}
\end{eqnarray}
In order to find the relevant Feynman rules we first expand potential in power series in $\Phi$. Differentiating the relation
\begin{eqnarray}
\rho (\rho (\Phi ))=\Phi \label{w12}
\end{eqnarray}
three times and puting $\Phi =0$\ we get 
\begin{subequations}
\label{w13}
\begin{gather}
\rho '(0)=-1 \\
3(\rho ''(0))^2+2\rho '''(0)=0.
\end{gather}
\end{subequations}
Assume that $\rho ''(0)\equiv \rho_2\neq 0$. We have then
\begin{eqnarray}
L=\frac{1}{2}\partial_{\mu}\Phi \partial^{\mu}\Phi-\frac{m^2}{2}\Phi^2-\frac{\lambda}{3!}\Phi^3-
\frac{5\frac{\lambda^2}{3m^2}}{4!}\Phi^4+... \label{w14}
\end{eqnarray}
where $m^2=8\alpha ,\;\lambda=-12\alpha \rho_2$\ and dots denote higher-order terms. Due to $\lambda \neq 0$\ the lowest
apriori nontrivial amplitude is $2\rightarrow 2$. The relevant graphs are shown on Fig.1

\begin{equation*}
\parbox{24mm}{
\begin{fmfgraph*}(51,48)
\fmfleftn{i}{2}\fmfrightn{o}{2}
\fmf{plain}{i1,o2}
\fmf{plain}{i2,o1}
\fmflabel{$1$}{i1}
\fmflabel{$2$}{i2}
\fmflabel{$3$}{o1}
\fmflabel{$4$}{o2}
\end{fmfgraph*}
}
+\;\;
\parbox{24mm}{
\begin{fmfgraph*}(51,48)
\fmfleftn{i}{2}\fmfrightn{o}{2}
\fmfforce{.3w,.5h}{x1}
\fmfforce{.7w,.5h}{x2}
\fmf{plain}{i2,x1}
\fmf{plain}{i1,x1}
\fmf{plain}{x1,x2}
\fmf{plain}{x2,o1}
\fmf{plain}{x2,o2}
\fmflabel{$1$}{i1}
\fmflabel{$2$}{i2}
\fmflabel{$3$}{o1}
\fmflabel{$4$}{o2}
\end{fmfgraph*}
}
+
\parbox{24mm}{
\begin{fmfgraph*}(51,38)
\fmfforce{.2w,1.15h}{i1}
\fmfforce{.8w,1.15h}{i2}
\fmfforce{.2w,-.15h}{o1}
\fmfforce{.8w,-.15h}{o2}
\fmfforce{.5w,.8h}{x1}
\fmfforce{.5w,.2h}{x2}
\fmf{plain}{i2,x1}
\fmf{plain}{i1,x1}
\fmf{plain}{x1,x2}
\fmf{plain}{x2,o1}
\fmf{plain}{x2,o2}
\fmflabel{$2$}{i1}
\fmflabel{$4$}{i2}
\fmflabel{$1$}{o1}
\fmflabel{$3$}{o2}
\end{fmfgraph*}
}
+\;\;
\parbox{24mm}{
\begin{fmfgraph*}(51,32)
\fmfforce{0w,1.3h}{i2}
\fmfforce{0w,-0.3h}{i1}
\fmfforce{.35w,.9h}{x1}
\fmfforce{.35w,.1h}{x2}
\fmfforce{1.12w,1.1h}{x3}
\fmfforce{1.12w,-.1h}{x4}
\fmfforce{.6w,.45h}{x5}
\fmfforce{.73w,.55h}{x6}
\fmf{plain}{i2,x1}
\fmf{plain}{i1,x2}
\fmf{plain}{x1,x2}
\fmf{plain}{x2,x5}
\fmf{plain}{x1,x4}
\fmf{plain}{x6,x3}
\fmf{plain,left=0.7}{x5,x6}
\fmflabel{$1$}{i1}
\fmflabel{$2$}{i2}
\fmflabel{$3$}{x4}
\fmflabel{$4$}{x3}
\end{fmfgraph*}
}
\end{equation*}
\begin{equation*}
\end{equation*}
\;\;\;\;\;\;\;\;\;\;\;\;\;\;\;\;\;\;\;\;\;\;\;\;\;\;\;\;\;\;\;\;\;\;\;\;\;\;\;\;\;\;\;\;\;\;\;\;\;\;\;\;\;\;\;Fig. 1\\
Using Feynman rules implied by (\ref{w14}) we immediately check that the contributions from these graphs sum to zero.

The results for $n\rightarrow n,\;n>2$, processes are ambigious for the same reasons as in Ref. \cite{b15}. If we consider
the amplitudes as calculated from $\Phi (t\mid \Phi_0)$\ by amputating the last external propagator, we obviously
obtain zero: there are no resonant pieces in the "external force" coming from lower order terms. On the other hand,
the  corrsponding Feynman graphs give ambigious contribution $0\over 0$. One should therefore consider the limit of vanishing
threemomenta in general amplitudes; however, this limit is also in general ambigious.

Consider now the general case when the first nontrivial amplitude is $n\rightarrow n$\ with some $n>2$. This corresponds
to $\rho_2=0$. We shall consider the most general case when first few derivatives of $\rho$\ vanish. Detailed analysis, 
based again on eq. (\ref{w12}) and given in Appendix, can be summarized as follows. Except $\rho'(0)=-1$, the first
nonvanishing derivative must be of even order, $\rho^{(2p)}(0)\neq 0$. Moreover, we arrive at the following conclusion:
\begin{eqnarray}
&&\rho^{(2k)}(0) \; \hbox{are arbitrary for}\;k=\;p,\;p+1,\;...,\;2p-1 \nonumber\\
&&\rho^{(2k+1)}(0)=0,\;k=\;p,\;p+1,\;...,\;2p-2 \label{w15} \\
&&{4p-1\choose2p}\ \left(\rho^{(2p)}(0) \right)^2+2\rho^{(4p-1)}(0)=0 \nonumber
\end{eqnarray}

Denote $\rho_n\equiv \rho^{(n)}(0)$; our lagrangian reads now 
\begin{eqnarray}
L=\frac{1}{2} \partial_{\mu}\Phi \partial^{\mu}\Phi -\alpha 
\left(2\Phi -\sum\limits_{k=p}^{2p-1}\frac{\rho_{2k}\Phi^{2k}}{(2k)!}
-\frac{\rho_{4p-1}}{(4p-1)!}\Phi^{4p-1}+...\right)^2\label{w16}
\end{eqnarray}
where, as usual, dots denote higher-order terms. Taking the square on {\bf RHS} of (\ref{w16}) and inspecting all terms 
carefully we conclude that the lowest nontrivial amplitude is $n\rightarrow n$\ with $n=2p$. Skiping all vertices which
are irrelevant for this process and using the last relation (\ref{w15}) we get 
\begin{eqnarray}
L=\frac{1}{2}\partial_{\mu}\Phi \partial^{\mu}\Phi -\frac{1}{2}m^2\Phi^2-\frac{\lambda}{(2p+1)!}\Phi^{2p+1}-
\frac{\frac{\lambda^2{4p+1\choose2p}\ }{2m^2(2p+1)}}{(4p)!}\Phi^{4p}\label{w17}
\end{eqnarray}
The graphs contributing to the $2p\rightarrow 2p$\ process are shown on Figs. 2 and 3.
\newpage
\begin{equation}
2p \left\{
\parbox{24mm}{
\begin{fmfgraph}(21mm,18mm)
\fmfleftn{i}{7}\fmfrightn{o}{7}
\fmfforce{.5w,.5h}{x}
\fmf{plain}{i1,o7}
\fmf{plain}{i7,o1}
\fmfv{decor.shape=square,decor.filled=full,decor.size=2thick}{x}
\fmfv{decor.shape=circ,decor.filled=full,decor.size=0.5thick}{o2,o3,o4,o5,o6}
\fmfv{decor.shape=circ,decor.filled=full,decor.size=0.5thick}{i2,i3,i4,i5,i6}
\end{fmfgraph}
}
\right\}2p
\nonumber
\end{equation}
\;\;\;\;\;\;\;\;\;\;\;\;\;\;\;\;\;\;\;\;\;\;\;\;\;\;\;\;\;\;\;\;\;\;\;\;\;\;\;\;\;\;\;\;\;\;\;\;\;\;\;\;\;\;\;Fig. 2
\begin{equation}
{}\;\;\;\;\;{}l \left\{
\parbox{24mm}{
\begin{fmfgraph}(21mm,18mm)
\fmfleftn{i}{7}\fmfrightn{o}{7}
\fmfforce{.5w,.5h}{x}
\fmfforce{.5w,-.5h}{y}
\fmf{plain}{i1,o7}
\fmf{plain}{x,y}
\fmf{plain}{i7,o1}
\fmfv{decor.shape=square,decor.filled=full,decor.size=2thick}{x}
\fmfv{decor.shape=circ,decor.filled=full,decor.size=0.5thick}{o2,o3,o4,o5,o6}
\fmfv{decor.shape=circ,decor.filled=full,decor.size=0.5thick}{i2,i3,i4,i5,i6}
\end{fmfgraph}
}
\right\}2p-l\;\;\;\;\;\;\;\;\;\;\;\;\;\;\;\;\;\;\;\;\;\;\;\;\;\;\;\;\;\;\;\;\;\;\;\;\;\;\;\;\;\; \nonumber
\end{equation}
\begin{equation*}
\;\;\;\;\;\;\;\;\;\;\;\;\;\;\;\;\;\;\;\;\;\;\;\;\;\;\;l=0,\;...,\;2p
\end{equation*}
\begin{equation}
2p-l \left\{
\parbox{24mm}{
\begin{fmfgraph}(21mm,18mm)
\fmfleftn{a}{7}\fmfrightn{b}{7}
\fmfforce{.5w,.5h}{y}
\fmfforce{.5w,1.5h}{x}
\fmf{plain}{a1,b7}
\fmf{plain}{x,y}
\fmf{plain}{a7,b1}
\fmfv{decor.shape=square,decor.filled=full,decor.size=2thick}{y}
\fmfv{decor.shape=circ,decor.filled=full,decor.size=0.5thick}{a2,a3,a4,a5,a6}
\fmfv{decor.shape=circ,decor.filled=full,decor.size=0.5thick}{b2,b3,b4,b5,b6}
\end{fmfgraph}
}
\right\}l\;\;\;\;\;\;\;\;\;\;\;\;\;\;\;\;\;\;\;\;\;\;\;\;\;\;\;\;\;\;\;\;\;\;\;\;\;\;\;\;\;\;\;\;\;\;\;\;\;\;\;\;\;\nonumber
\end{equation}
\;\;\;\;\;\;\;\;\;\;\;\;\;\;\;\;\;\;\;\;\;\;\;\;\;\;\;\;\;\;\;\;\;\;\;\;\;\;\;\;\;\;\;\;\;\;\;\;\;\;\;\;\;\;\;Fig. 3

The total contribution coming from these graphs is readily found to be
\begin{eqnarray}
\frac{-i\lambda^2}{2m^2}\left[\sum\limits_{l=0}^{2p}{2p\choose l}\ {2p\choose2p-l}\ \frac{1}{(2p-2l)^2-1}+
\frac{1}{(2p+1)}{4p+1\choose2p}\ \right] \nonumber
\end{eqnarray}
However, as it is shown in Appendix, the expression in square brackets vanishes.

The explicit construction of arbitrary function $\rho$\ is given in Refs. \cite{b21}, \cite{b22}. Using the results 
contained there we can define infinity of models sharing the property of having vanishing tree-level threshold amplitudes.
\newpage
{\bf \Huge Appendix}

Let $\rho$\ and $\sigma$\ be smooth functions of one real variable and
\begin{equation*}
f=\sigma \circ \rho
\end{equation*}
We wish to find $n$-th derivative of $f$. Its general structure reads
\begin{eqnarray}
f^{(n)}(\Phi )=\sum\limits_{k=1}^n\sigma^{(k)}(\rho (\Phi ))\cdot F_k^n(\rho '(\Phi ),\;...) \label{w18}
\end{eqnarray}
where $F_k^n$\ are polynomial functions of $\rho '$\ and higher derivatives of $\rho $\ up the order $n-k+1$\ (see below). 
Taking derivative of (\ref{w18}) one arrives at the following reccurence relations
\begin{eqnarray}
&&F_k^{n+1}=\rho 'F_{k-1}^n+(F_k^n)',\;\;2\leq k\leq n\nonumber \\
&&F_{n+1}^{k+1}=\rho 'F_n^n \label{w19} \\
&&F_1^{n+1}=(F_1^n)' \nonumber
\end{eqnarray}
The solution to (\ref{w19}) can be written as
\begin{eqnarray}
F_k^n=\frac{1}{k!}\frac{d^2(\rho^k)}{d\Phi^n}\mid_{\rho =0}\label{w20}
\end{eqnarray}
The notation here is as follows: we take $n$-th derivative of $\rho^k$\ and neglect all terms containing at least one 
factor $\rho$\ with no derivatives. To prove (\ref{w20}) let us note the following identity
\begin{eqnarray}
\frac{d^n(\rho^k)}{d\Phi^n}=\frac{d^n(\rho^k)}{d\Phi^n}\mid_{\rho =0}+k\rho \frac{d^n(\rho^{k-1})}{d\Phi^n}\mid_{\rho =0}
+... \label{w21}
\end{eqnarray}
where the dots denote terms containing $\rho^2$\ and higher powers of $\rho$. Differentiating again (\ref{w21}) and
neglecting terms containing $\rho$\ we obtain (\ref{w19}).

Let us apply this in the case $\sigma =\rho$\ with $\rho$\ as in the main body of the paper and $\Phi =0$. We know already
that $\rho '(0)=-1$. Moreover,
\begin{eqnarray}
(\rho \circ \rho )^{(n)}(0)=\sum\limits_{k=1}^n\rho^{(k)}(0)F_k^n(\rho '(0),\;...) \label{w22}
\end{eqnarray}
Assume that $\rho^{(k)}(0)=0$\ for $2<\;k\leq \;l,\;\rho^{(l+1)}(0) \neq 0$. We show that $l=2p-1$; indeed, assume $l=2p-2$;
then, due to $(\rho \circ \rho )^{(n)}=\delta_{n1}$, (\ref{w22}) implies
\begin{eqnarray}
\rho^{(2p-1)}(0)\cdot (\rho '(0))^{2p-1}+\rho '(0)\rho^{(2p-1)}(0)+\sum\limits_{k=2}^{2p-2}\rho^{(k)}(0)F_k^n=0\label{w23}
\end{eqnarray}
Now, $\rho '(0)=-1$\ and the last term on {\bf LHS} vanishes; consequently, $\rho^{(2p-1)}(0)=0$, contrary to the assunption.
So $\rho^{(k)}(0)=0,\;2<\;k\leq\;2p-1,\;\rho^{(2p)}(0)\neq 0$. Let us take now $2p\leq \; n\leq \; 4p-2$; then
\begin{eqnarray}
\rho^{(n)}(0)((-1)^n-1)+\sum\limits_{k=2}^{n-1}\rho^{(k)}(0)F_k^n=0\label{w24}
\end{eqnarray}
Consider the last term on {\bf LHS}. Due to the assumption made above the sum starts effectively from $k=2p$. But $F_k^n=0$\
for $k\geq \; 2p,\;n\leq \; 4p-2$; indeed, (\ref{w20}) implies that the maximal order of derivatives of $\rho$\ entering 
$F_k^n$\ is $n-k+1\leq \;2p-1$; moreover, for $k\leq \;n-1$\ each term entering $F_k^n$\ contains higher than first derivative of $\rho$.

Finally, take $n=4p-1$; we get 
\begin{eqnarray}
-2\rho^{(4p-1)}(0)+\sum\limits_{k=2}^{4p-2}\rho^{(k)}(0)F_k^{4p-1}=0\label{w25}
\end{eqnarray}
The only term in the sum on the {\bf LHS} which is nonvanishing corresponds to $k=2p$. Let us calculate
\begin{eqnarray}
F_{2p}^{4p-1}=\frac{1}{(2p)!}\frac{d^{4p-1}(\rho^{2p})}{d\Phi^{4p-1}}\mid_{\rho =0}\label{w26}
\end{eqnarray}
The only terms contributing to the {\bf RHS} are those proportional to $(\rho ')^{2p-1}\cdot \rho^{(2p)}=-\rho^{(2p)}$. It 
is easy to see that the total coefficient in front of this term is ${4p+1\choose 2p}$\ which, together with (\ref{w25}), 
proves the last identity (\ref{w15}).

Finally, we shall prove the identity
\begin{eqnarray}
\sum\limits_{l=0}^{2p}{2p\choose l}{2p\choose 2p-l}\frac{1}{(2p-2l)^2-1}+\frac{1}{(2p+1)}{4p+1\choose 2p}=0\label{w27}
\end{eqnarray}
We have
\begin{eqnarray}
&&\sum\limits_{l=0}^{2p}{2p\choose l}{2p\choose 2p-l}\frac{1}{(2p-2l)^2-1}=\frac{1}{2}\sum\limits_{l=0}^{2p}
{2p\choose l}{2p\choose 2p-l}\left(\frac{1}{2p-2l-1}-\frac{1}{2p-2l+1}\right) \nonumber \\
&&=-\sum\limits_{l=0}^{2p}{2p\choose l}{2p\choose 2p-l}\frac{1}{2p-2l+1}\label{w28}
\end{eqnarray}
where the last equality results from the change of summation variable $l\rightarrow \; 2p-l$\ in the first term of the 
expression in the middle.

Consider the identity
\begin{eqnarray}
(1+x)^{2p}(1+y)^{2p}=\sum\limits_{k,\;l=0}^{2p}{2p\choose l}{2p\choose k}x^ly^k;\nonumber
\end{eqnarray}
integrating with respect to $x$\ from $0$\ to $x$, putting $y=x$\ and comparying the coefficients in front of 
$x^{2p+1}$\ we obtain
\begin{eqnarray}
\sum\limits_{l=0}^{2p}{2p\choose l}{2p\choose 2p-l}\frac{1}{2p-l+1}=\frac{1}{2p+1}{4p+1\choose 2p} \label{w29}
\end{eqnarray}
On the other hand
\begin{eqnarray}
&&\sum\limits_{l=0}^{2p}{2p\choose l}{2p\choose 2p-l}\left(\frac{1}{2p-l+1}-\frac{1}{2p-2l+1}\right)=-
\sum\limits_{l=0}^{2p}{2p\choose l-1}{2p\choose 2p-l}\frac{1}{2p-2l+1}= \nonumber \\
&&=-\sum\limits_{k+l=2p-1}{2p\choose l}{2p\choose k}\frac{1}{k-l}=0;\label{w30}
\end{eqnarray}
(\ref{w27}) follows now easily from (\ref{w28})$\div$ (\ref{w30}).

\end{fmffile}


\begin{thebibliography}{99}
\bibitem{b1}
M. Libanov, V. Rubakov, S. Troitsky, Part. Nucl. Phys. \underline{\bf 28}, 551 (1997) (in Russian)
\bibitem{b2}
M. Voloshin, Nucl. Phys. \underline{B383}, 233 (1992)
\bibitem{b3}
E. Argyres, R. Kleiss, C. Papadopoulos, Nucl. Phys. \underline{\bf 391}, 42, 57 (1993)
\bibitem{b4}
L. Brown, Phys. Rev. \underline{\bf D46}, 4125 (1992)
\bibitem{b5}
B. Smith, Phys. Rev. \underline{\bf D47}, 3521 (1993)
\bibitem{b6}
L. Brown, Cheng-Xing Zhai, Phys. Rev. \underline{\bf D47}, 5526 (1993)
\bibitem{b7}
M. Libanov, V. Rubakov, S. Troitsky, Nucl. Phys. \underline{\bf B412}, 607 (1994)
\bibitem{b8}
E. Argyres, R. Kleiss, C. Papadopoulos, Nucl. Phys. \underline{\bf B395}, 3 (1993)
\bibitem{b9}
M. Voloshin, Phys. Rev. \underline{\bf D47}, 357 (1993)
\bibitem{b10}
B. Smith, Phys. Rev. \underline{\bf D47}, 3518 (1993)
\bibitem{b11} M. Voloshin, Phys. Rev. \underline{\bf D47}, 2573 (1993)

B. Smith, Phys. Rev. \underline{\bf D49}, 1081 (1994)

E. Argyres, R. Kleiss, C. Papadopoulos, Phys. Lett. \underline{\bf B308}, 315 (1993)
\bibitem{b12}
See first reference in [11]
\bibitem{b13}
B. Smith, Phys. Rev. \underline{\bf D47}, 5531 (1993)

M. Libanov, D. Son, S. Troitsky, Phys. Rev. \underline{\bf D52}, 3679 (1995)
\bibitem{b14}
M. Libanov, V. Rubakov, S. Troitsky, INR pre-print INR 823/93
\bibitem{b15}
M. Libanov, V. Rubakov, S. Troitsky, Phys. Lett. \underline{\bf B318}, 134 (1993)
\bibitem{b16}
J. Domienik, J. Gonera, P. Kosi\'nski, Acta Phys. Polon \underline{\bf B32}, 2813 (2001)
\bibitem{b17}
J. Gonera, Phys. Rev.  \underline{\bf D66}, 105010 (2002)
\bibitem{b18}
J. Domienik, P. Kosi\'nski, to be published
\bibitem{b19}
S. Weinberg, Physica \underline{\bf 96A} (1979), 327

S. Weinberg, The Quantum Theory of Fields II, Cambridge University Press, 1996
\bibitem{b20}
L. Landau, E. Lifshic, Mechanics, Pergamon Press 1963
\bibitem{b21}
C. Gonera, P. Kosi\'nski, P. Ma\'slanka, Phys. Lett. A \underline{\bf 289}, 66 (2001)
\bibitem{b22}
C. Gonera, More about generalized maximally superintegrable systems of Winternitz type, hep-th/0207182
\end{thebibliography}
\end{document}